\documentclass[10pt,letterpaper]{scrartcl}
\usepackage[utf8]{inputenc}
\usepackage[english]{babel} 
\usepackage{gensymb}
\usepackage[paper=letterpaper,left=25mm,right=25mm,top=25mm,bottom=30mm]{geometry}
\usepackage[numbers,sort&compress,merge,round]{natbib}
\usepackage{dblfloatfix}
\usepackage{float}
\usepackage[breaklinks, colorlinks=true, citecolor=black, urlcolor=black, linkcolor=black]{hyperref}

\usepackage{authblk}
\usepackage{multicol}
\usepackage{graphicx}

\usepackage{amsmath}
\usepackage[format=plain,labelfont=bf,up,textfont=it,up]{caption}
\usepackage{sidecap}

\usepackage[bottom, hang]{footmisc}
\usepackage{scrpage2}
\usepackage[dvipsnames]{xcolor}
\usepackage{graphicx}
\definecolor{boxcolor}{HTML}{e3e9f2}
\usepackage{nameref}

\title{\vspace{-1.0cm}\Large Deep learning to represent sub-grid processes in climate models}

\author{\large Stephan Rasp\thanks{Meteorological Institute, Ludwig-Maximilian-University, Munich, Germany. Corresponding author: s.rasp@lmu.de}\ \textsuperscript{$\dagger$} \, \, Michael S. Pritchard\footnote{Department of Earth System Science, University of California, Irvine, CA, USA}\, \, \, Pierre Gentine\thanks{Columbia University, Department of Earth and Environmental Engineering, Earth Institute, and Data Science Institute, New York, NY, USA}}
\date{\vspace{-5ex}}
\begin{document}

\maketitle

\begin{figure}[bp]
\noindent\colorbox{boxcolor}{%
	\parbox{\textwidth}{%
\textbf{\large Plain language summary}
Current climate models are too coarse to resolve many of the atmosphere's most important processes. Traditionally, these sub-grid processes are heuristically approximated in so-called parameterizations. However, imperfections in these parameterizations, especially for clouds, have impeded progress towards more accurate climate predictions for decades. Cloud-resolving models alleviate many of the gravest issues of their coarse counterparts but will remain too computationally demanding for climate change predictions for the foreseeable future. Here we use deep learning to leverage the power of short-term cloud-resolving simulations for climate modeling. Our data-driven model is fast and accurate thereby showing the potential of novel, machine learning-based approaches to climate model development.
    }%
}
\end{figure}

\begin{multicols}{2}
%\subsection*{Abstract}
\noindent~\textbf{
The representation of nonlinear sub-grid processes, especially clouds, has been a major source of uncertainty in climate models for decades. Cloud-resolving models better represent many of these processes and can now be run globally but only for short-term simulations of at most a few years because of computational limitations. Here we demonstrate that deep learning can be used to capture many advantages of cloud-resolving modeling at a fraction of the computational cost. We train a deep neural network to represent all atmospheric sub-grid processes in a climate model by learning from a multi-scale model in which convection is treated explicitly. The trained neural network then replaces the traditional sub-grid parameterizations in a global general circulation model in which it freely interacts with the resolved dynamics and the surface-flux scheme. The prognostic multi-year simulations are stable and closely reproduce not only the mean climate of the cloud-resolving simulation but also key aspects of variability, including precipitation extremes and the equatorial wave spectrum. Furthermore, the neural network approximately conserves energy despite not being explicitly instructed to. Finally, we show that the neural network parameterization generalizes to new surface forcing patterns but struggles to cope with temperatures far outside its training manifold. Our results show the feasibility of using deep learning for climate model parameterization. In a broader context, we anticipate that data-driven Earth System Model development could play a key role in reducing climate prediction uncertainty in the coming decade.
}

%\subsection*{Introduction}
\noindent Many of the atmosphere's most important processes occur on scales smaller than the grid resolution of current climate models, around 50--100 km horizontally. Clouds, for example, can be as small as a few hundred meters; yet they play a crucial role in determining the earth's climate by transporting heat and moisture, reflecting and absorbing radiation, and producing rain. Climate change simulations at such fine resolutions are still many decades away \cite{Schneider2017}. To represent the effects of such sub-grid processes on the resolved scales, physical approximations---called \textit{parameterizations}---have been heuristically developed and tuned to observations over the last decades \cite{ Hourdin2017}. However, owing to the sheer complexity of the underlying physical system, significant inaccuracies persist in the parameterization of clouds and their interaction with other processes, such as boundary-layer turbulence and radiation \cite{Stevens2013, Bony2015, Schneider2017}. These inaccuracies manifest themselves in stubborn model biases \cite{Oueslati2015, Arnold2015a, Gentine2013a} and large uncertainties about how much the earth will warm as a response to increased greenhouse gas concentrations \cite{Bony2005, Sherwood2014, Schneider2017}. To improve climate predictions, therefore, novel, objective and computationally efficient approaches to sub-grid parameterization development are urgently needed.

Cloud-resolving models (CRMs) alleviate many of the issues related to parameterized convection. At horizontal resolutions of at least 4 km deep convection can be explicitly treated \cite{Weisman1997}, which substantially improves the representation of land-atmosphere coupling \cite{Sun2016a, Leutwyler2017}, convective organization \cite{Muller2015} and weather extremes. Further increasing the resolution to a few hundred meters allows for the direct representation of the most important boundary-layer eddies, which form shallow cumuli and stratocumuli. These low clouds are crucial for the Earth's energy balance and the cloud-radiation feedback \cite{Soden2011}. CRMs come with their own set of tuning and parameterization decisions but the advantages over coarser models are substantial. Unfortunately, global CRMs will be too computationally expensive for climate change simulations for many decades \cite{Schneider2017}. Short-range simulations covering periods of months or even a few years, however, are beginning to be feasible and are in development at modeling centers around the world \cite{Miyamoto2013, Bretherton2015, Yashiro2016, Klocke2017}. 

In this study, we explore whether deep learning can provide an objective, data-driven approach to utilize high-resolution modeling data for climate model parameterization. The paradigm shift from heuristic reasoning to machine learning has transformed computer vision and natural language processing over the last few years \cite{LeCun2015} and is starting to impact more traditional fields of science. The basic building blocks of deep learning are deep neural networks which consist of several inter-connected layers of nonlinear nodes \cite{Goodfellow2016}. They are capable of approximating arbitrary nonlinear functions \cite{Nielsen2015} and can easily be adapted to novel problems. Furthermore, they can handle large datasets during training and provide fast predictions at inference time. All of these traits make deep learning an attractive approach for the problem of sub-grid parameterization.

Extending on previous offline or single-column neural network cumulus parameterization studies \cite{Krasnopolsky2013, Brenowitz2018, Gentine2018}, here we take the essential step of implementing the trained neural network in a global climate model and running a stable, prognostic multi-year simulation. To show the potential of this approach we compare key climate statistics between the deep learning-powered model and its training simulation. Furthermore, we tackle two crucial questions for a climate model implementation: first, does the neural network parameterization conserve energy; and second, to what degree can the network generalize outside of its training climate? We conclude by highlighting crucial challenges for future data-driven parameterization development.

\subsection*{Model and neural network setup}
Our base model is the super-parameterized Community Atmosphere Model v3.0 (SPCAM) \cite{Collins2006} in an aquaplanet setup (see \nameref{sec:MM} for details). The sea surface temperatures (SSTs) are fixed and zonally invariant with a realistic equator-to-pole gradient \cite{Andersen2012}. The model has a full diurnal cycle but no seasonal variation. The horizontal grid spacing of the global circulation model (GCM) is approximately 2 degrees with 30 vertical levels. The GCM time step is 30 minutes. In super-parameterization, a two-dimensional CRM is embedded in each global circulation model grid column \cite{Khairoutdinov2001}. This CRM explicitly resolves deep convective clouds and includes parameterizations for small-scale turbulence and cloud microphysics. In our setup, we use eight 4 km-wide columns with a CRM time step of 20 seconds, after Ref.~\cite{Pritchard2014}. For comparison, we also run a simulation with the traditional parameterization suite (CTRLCAM) that is based on an undilute plume parameterization of moist convection. CTRLCAM exhibits many typical problems associated with traditional sub-grid cloud parameterizations: a double inter-tropical convergence zone (ITCZ) \cite{Oueslati2015}; too much drizzle and missing precipitation extremes; and an unrealistic equatorial wave spectrum with a missing Madden-Julian-Oscillation (MJO). In contrast, SPCAM captures the key benefits of full three-dimensional CRMs in improving the realism all of these issues with respect to observations \cite{Benedict2009, Arnold2015, Kooperman2018}. In this context, a key test for a neural network parameterization is whether it learns sufficiently from the explicitly resolved convection in SPCAM to remedy such problems while being computationally more affordable.

\begin{figure*}[bh!]
    \centering
    \includegraphics[width=1\linewidth]{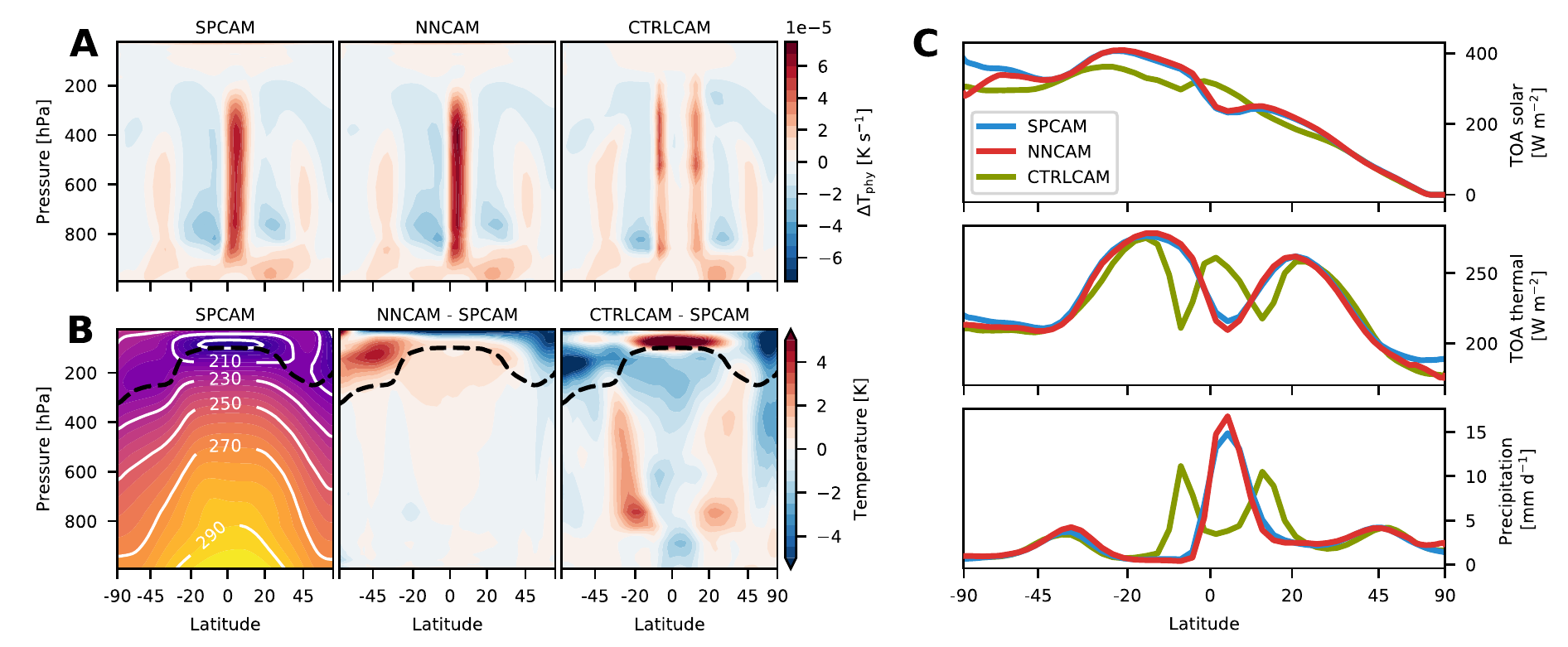}
    \caption{All figures show longitudinal and five year-temporal averages. (A) Mean convective and radiative sub-grid heating rates $\Delta T_{\mathrm{phy}}$. (B) Mean temperature $T$ of SPCAM and biases of NNCAM and CTRLCAM relative to SPCAM. The dashed black line denotes the approximate position of the tropopause, determined by a $\partial p \theta$ contour. (C) Mean shortwave (solar) and longwave (thermal) net fluxes at the top of the atmosphere and precipitation. Note that in all figures the latitude axis is area-weighted.}
    \label{fig:1}
\end{figure*}

Analogous to a traditional parameterization, the task of the neural network is to predict the sub-grid tendencies as a function of the atmospheric state at every time step and grid column (Table \ref{tab:vars}). Specifically, we selected the following input variables: the temperature $T(z)$, specific humidity $Q(z)$ and wind profiles $V(z)$, surface pressure $P_s$, incoming solar radiation $S_{\mathrm{in}}$ and the sensible $H$ and latent heat fluxes $E$. These variables mirror the information received by the CRM and radiation scheme with a few omissions (\nameref{sec:MM}). The output variables are: the sum of the CRM and radiative heating rates $\Delta T_{\mathrm{phy}}$, the CRM moistening rate $\Delta Q_{\mathrm{phy}}$, the net radiative fluxes at the top of atmosphere and surface $F_{\mathrm{rad}}$ and precipitation $P$. The input and output variables are stacked to vectors $\mathbf{x} = [T(z), Q(z), V(z), P_s, S_{\mathrm{in}}, H, E]^T$ with length 94 and $\mathbf{y} = [\Delta T_{\mathrm{phy}}(z), \Delta Q_{\mathrm{phy}}(z), F_{\mathrm{rad}}, P]^T$ with length 65 and normalized to have similar orders of magnitude (\nameref{sec:MM}). We omit condensed water to reduce the complexity of the problem (see \nameref{sec:discussion}). Furthermore, there is no momentum transport in our version of SPCAM. Informed by our previous sensitivity tests \cite{Gentine2018} we use one year of SPCAM simulation as training data for the neural network, amounting to around 140 million training samples.  

The neural network itself $\mathbf{\hat{y}} = \mathcal{N}(\mathbf{x})$ is a nine layer deep, fully-connected network with 256 nodes in each layer. In total, the network has around half a million parameters that are optimized to minimize the mean squared error between the network's predictions $\mathbf{\hat{y}}$ and the training targets $\mathbf{y}$ (see \nameref{sec:MM}). This neural network architecture is informed by our previous sensitivity tests \cite{Gentine2018}. Using deep rather than shallow networks has two main advantages: first, deeper, larger networks achieve lower training losses; and second, deep networks proved more stable in the prognostic simulations (for details see \nameref{sec:MM} and Fig.~\ref{fig:S1}). Unstable modes and unrealistic artifacts have been the main issue in previous studies that used shallow architectures \cite{Krasnopolsky2013, Brenowitz2018}.

Once trained, the neural network replaces the super-parameterization's CRM as well as the radiation scheme in CAM (NNCAM). In our prognostic global simulations, the neural network parameterization interacts freely with the resolved dynamics as well as with the surface flux scheme. The neural network parameterization speeds up the model significantly: NNCAM's physical parameterization is around 20 times faster than SPCAM's and even 8 times faster than NNCAM's, in which the radiation scheme is particularly expensive. The key fact to keep in mind is that the neural network does not become more expensive at prediction time even when trained with higher-resolution training data. The approach laid out here should, therefore, scale easily to neural networks trained with vastly more expensive three-dimensional global CRM simulations.

The subsequent analyses are computed from five-year prognostic simulations after a one-year spin-up. All neural network, model and analysis code is available online (\nameref{sec:MM}).

\subsection*{Results}
\subsubsection*{Mean climate}

\begin{figure*}[bh!]
    \centering
    \includegraphics[width=1\linewidth]{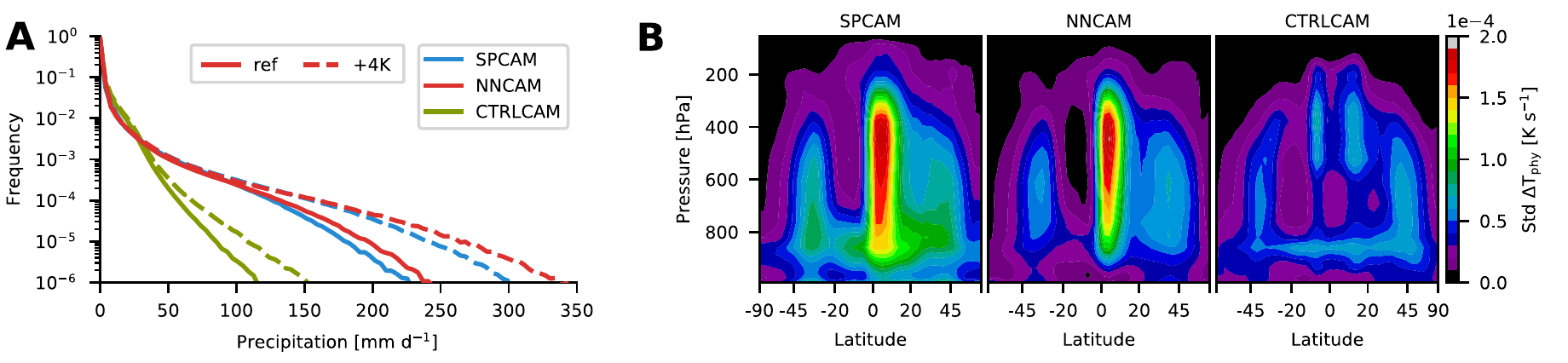}
    \caption{(A) Precipitation histogram of time-step (30 minutes) accumulation. The bin width is 3.9 mm d$^{-1}$. Solid lines denote simulations for reference SSTs. Dashed lines denote simulations for +4K SSTs (explanation in Generalization section). The neural network in the +4K case is NNCAM-ref+4K. (B) Zonally averaged temporal standard deviation of convective and radiative sub-grid heating rates $\Delta T_{\mathrm{phy}}$.}
    \label{fig:2}
\end{figure*}

To assess NNCAM's ability to reproduce SPCAM's climate we start by comparing the mean sub-grid tendencies and the resulting mean state. The mean sub-grid heating (Fig.~\ref{fig:1}\textit{A}) and moistening rates (Fig.~\ref{fig:S2}) of SPCAM and NNCAM are in close agreement with a single latent heating tower at the ITCZ and secondary free-tropospheric heating maxima at the mid-latitude storm tracks. The ITCZ peak, which is co-located with the maximum SSTs at 5\degree N, is slightly sharper in NNCAM compared to SPCAM. In contrast, CTRLCAM exhibits a double ITCZ signal, a common issue of traditional convection parameterizations \cite{Oueslati2015}.  The resulting mean state in temperature (Fig.~\ref{fig:1}\textit{B}), humidity and wind (Fig.~\ref{fig:S2}\textit{B} and \textit{C}) of NNCAM also closely resembles SPCAM throughout the troposphere. The only larger deviations are temperature biases in the stratosphere. Since the mean heating rate bias there is small, the temperature anomalies most likely have a secondary cause---for instance differences in circulation or internal variability. In any case, these deviations are not of obvious concern because the upper atmosphere is poorly resolved in our setup and highly sensitive to changes in the model setup (Fig.~\ref{fig:S5}\textit{C} and \textit{D}). In fact, CTRLCAM has even larger differences compared to SPCAM in the stratosphere but also throughout the troposphere for all variables.

The radiative fluxes predicted by the neural network parameterization also closely match those of SPCAM for most of the globe, whereas CTRLCAM has large differences in the tropics and subtropics caused by its double ITCZ bias (Figs.~\ref{fig:1}\textit{C} and \ref{fig:S2}\textit{D}). Towards the poles NNCAM's fluxes diverge slightly, the reasons for which are yet unclear. The mean precipitation of NNCAM and SPCAM follows the latent heating maxima with a peak at the ITCZ, which again is slightly sharper for NNCAM. 

In general, the neural network parameterization, freely interacting with the resolved dynamics, reproduces the most important aspects of its training model's mean climate to a remarkable degree, especially compared to the standard parameterization.

\subsubsection*{Variability}

Next, we investigate NNCAM's ability to capture SPCAM's higher-order statistics---a crucial test since climate modeling is as much concerned about variability as it is about the mean. One of the key statistics for end users is the precipitation distribution (Fig.~\ref{fig:2}\textit{A}). CTRLCAM shows the typical deficiencies of traditional convection parameterizations---too much drizzle and a lack of extremes. SPCAM remedies these biases and has been shown to better fit to observations \cite{Kooperman2018}. The precipitation distribution in NNCAM closely matches that of SPCAM, including the tail. The rarest events are slightly more common in NNCAM than in SPCAM, which is consistent with the narrower and stronger ITCZ (Fig.~\ref{fig:1}\textit{A} and \textit{C}). 

\begin{SCfigure*}[][th!]
	\centering
	\includegraphics[width=1.2\linewidth]{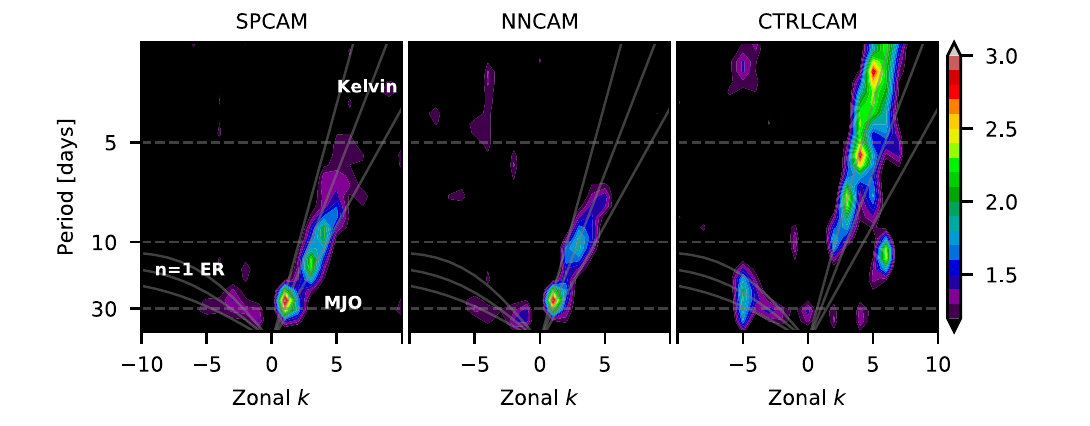}
	\caption{Space-time spectrum of the equatorially symmetric component of 15S-15N daily precipitation anomalies after Fig. 3b in Ref. \cite{Wheeler1999}. Negative (positive) values denote westward (eastward) traveling waves.}
	\label{fig:3}
\end{SCfigure*}

\begin{figure*}[bh!]
	\centering
	\includegraphics[width=1\linewidth]{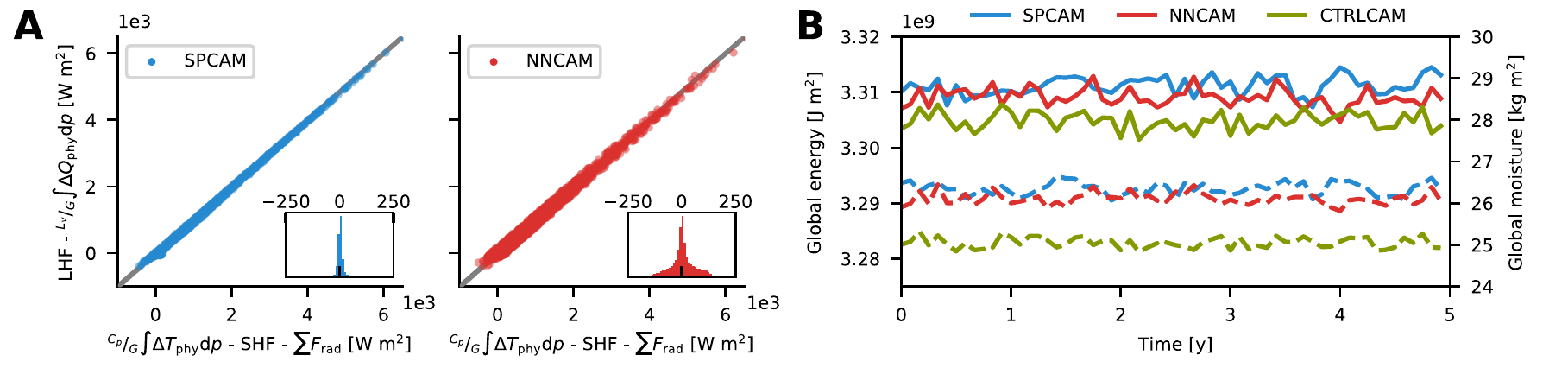}
	\caption{(A) Scatter plots of vertically integrated column heating $^{C_p}/_G \int \Delta T_{\mathrm{phy}} \mathrm{d}p$ minus the sensible heat flux $H$ and the sum of the radiative fluxes at the boundaries $\sum F_{\mathrm{rad}}$ against the vertically integrated column moistening $^{L_v}/_G \int \Delta T_{\mathrm{phy}} \mathrm{d}p$ minus the latent heat flux $H$. Each dot represent a single prediction at a single column. A total of ten time steps are shown. Inset show distribution of differences. (B) Globally integrated total energy (static, potential and kinetic; solid) and moisture (dashed) for the five-year simulations after one year of spin-up.}
	\label{fig:4}
\end{figure*}

We now focus on the variability of the heating and moistening rates (Figs.~\ref{fig:2}\textit{B} and \ref{fig:S3}\textit{A}). Here, NNCAM shows reduced variance compared to SPCAM and even CTRLCAM, mostly located at the shallow cloud level around 900 hPa and in the boundary-layer. Snapshots of instantaneous heating and moistening rates (Fig.~\ref{fig:S3}\textit{B} and \textit{C}) confirm that the neural network's predictions are much smoother, i.e. they lack the vertical and horizontal variability of SPCAM and CTRLCAM. We hypothesize that this has two separate causes: first, low training skill in the boundary-layer \cite{Gentine2018} suggests that much of SPCAM's variability in this region is chaotic and, therefore, has limited inherent predictability. Faced with such seemingly random targets during training, the deterministic neural network will opt to make predictions that are close to the mean in order to lower its cost function across samples. Second, the omission of condensed water in our network inputs and outputs limits NNCAM's ability to produce sharp radiative heating gradients at the shallow cloud tops.
% the ability of NNCAM to produce sharp radiative heating gradients at the shallow cloud tops. Perhaps counter-intuitively, the offline fits show that the heating rates associated with low clouds are well captured despite the lack of liquid and ice water \cite{Gentine2018}. This suggests that the neural network learns to infer the location of clouds from gradients in the temperature and water vapor profiles. In the prognostic implementation, however, the cloud top heating and cooling are largely missing (Fig.~S3\textit{B} and \textit{C}), most likely because NNCAM is not able to produce such gradients itself. Taking condensed water into account may fix this issue but adds non-trivial complexity (see Discussion). 
Because the circulation is mostly driven by mid-tropospheric heating in tropical deep convection and mid-latitude storms, however, the lack of low-tropospheric variability does not seem to negatively impact the mean state and precipitation predictions. This result is also of interest for climate prediction in general.

\begin{figure*}[b]
	\centering
	\includegraphics[width=1\linewidth]{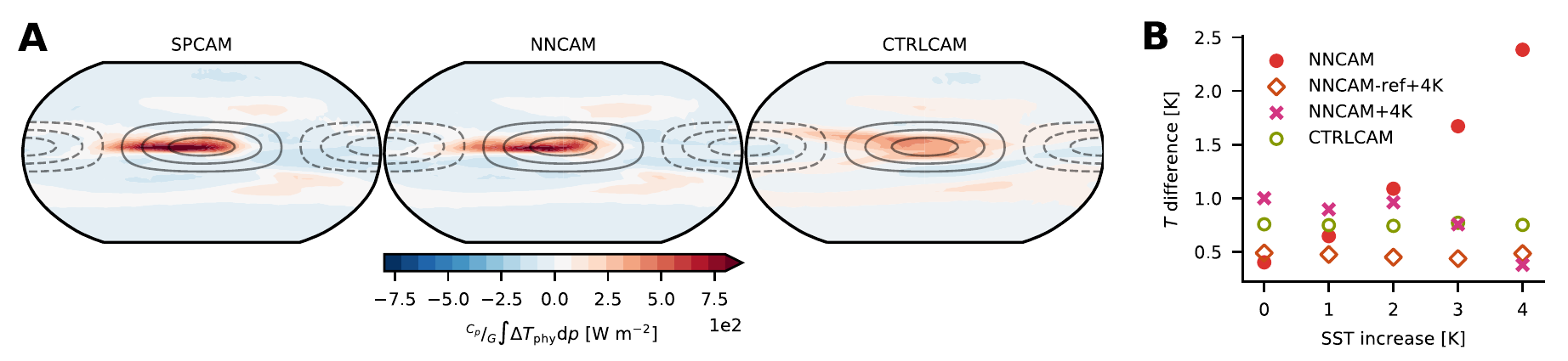}
	\caption{(A) Vertically integrated mean heating rate $^{C_p}/_G \int \Delta T_{\mathrm{phy}}  \mathrm{d}p$ for zonally perturbed SSTs. Contour lines show SST perturbation in 1 K intervals starting at 0.5 K. Dashed contours represent negative values. (B) Global mean mass-weighted absolute temperature difference relative to SPCAM reference at each SST increment. The different NNCAM experiments are explained in the corresponding text.}
	\label{fig:5}
\end{figure*}

The tropical wave spectrum \cite{Wheeler1999} depends vitally on the interplay between convective heating and large-scale dynamics. This makes it a demanding, indirect test of the neural network parameterization's ability to interact with the dynamical core. Current-generation climate models are still plagued by issues in representing tropical variability: in CTRLCAM, for instance, moist Kelvin waves are too active and propagate too fast while the MJO is largely missing (Fig.~\ref{fig:3}). SPCAM drastically improves the realism of the wave spectrum \cite{Benedict2009}, including in our aquaplanet setup \cite{Andersen2012}. NNCAM captures the key improvements of SPCAM relative to CTRLCAM: a damped Kelvin wave spectrum, albeit slightly weaker and faster in NNCAM, and an MJO-like intra-seasonal, eastward traveling disturbance. The background spectra also agree well with these results (Fig.~\ref{fig:S6}\textit{A})

Overall, NNCAM's ability to capture key advantages of the cloud-resolving training model---representing precipitation extremes and producing realistic tropical waves---is to some extent unexpected and represents a major advantage compared to traditional parameterizations.

\subsubsection*{Energy conservation}

A necessary property of any climate model parameterization is that it conserves energy. In our setup, energy conservation is not prescribed during network training. Despite this, NNCAM conserves column moist static energy to a remarkable degree (Fig.~\ref{fig:4}\textit{A}). Note that because of our omission of condensed water, the balance shown is only approximately true and exhibits some scatter even for SPCAM. The spread is slightly larger for NNCAM, but all points lie within a reasonable range, which shows that NNCAM never severely violates energy conservation. These results suggest that the neural network has approximately learned the physical relation between the input and output variables without being instructed to. This permits a simple post-processing of the neural network's raw predictions to enforce exact energy conservation. We tested this correction without noticeable changes to the main results. Conservation of total moisture is equally as important but the lack of condensed water makes even an approximate version impossible. 

The globally integrated total energy and moisture are also stable without noticeable drift or unreasonable scatter for multi-year simulations (Fig.~\ref{fig:4}\textit{B}). This is still true for a 50-year NNCAM simulation that we ran as a test. The energy conservation properties of the neural network parameterization are promising and show that, to a certain degree, neural networks can learn higher-level concepts and physical laws from the underlying dataset.

\subsubsection*{Generalization}

A key question for the prediction of future climates is whether such a neural network parameterization can generalize outside of its training manifold. To investigate this we run a set of sensitivity tests with perturbed SSTs. We begin by breaking the zonal symmetry of our reference state by adding a wavenumber one SST perturbation with 3K amplitude (Fig.~\ref{fig:5}\textit{A}; \nameref{sec:MM}). Under such a perturbation SPCAM develops a thermally direct Walker circulation within the tropics with convective activity concentrated at the downwind sector of the warm pool. The neural network trained with the zonally invariant reference SSTs only (NNCAM) is able to generate a similar heating pattern even though the heating maximum is slightly weaker and more spread out. The resulting mean temperature state in the troposphere is also in close agreement, with biases of less than 1 K (Fig.~\ref{fig:S4}). Moreover, NNCAM runs stably despite the fact that the introduced SST perturbations exceed the training climate by as much as 3 K. CTRLCAM, for comparison, has a drastically damped heating maximum and a double ITCZ to the west of the warm pool. 

Our next out-of-sample test is a global SST warming of up to 4 K in 1 K increments. We use the mass-weighted absolute temperature differences relative to the SPCAM reference solution at each SST increment as a proxy for the mean climate state difference (Fig.~\ref{fig:5}\textit{B}). The neural network trained with the reference climate only (NNCAM) is unable to generalize to much warmer climates. A look at the mean heating rates for the +4K SST simulation reveals that the ITCZ signal is washed out and unrealistic patterns develop in and above the boundary-layer (Fig.~\ref{fig:S5}\textit{B}). As a result the temperature bias is significant, particularly in the stratosphere (Fig.~\ref{fig:S5}\textit{D}). This suggests that the neural network cannot handle temperatures that exceed the ones seen during training. To test the opposite case, we also trained a neural network with data from the +4K SST SPCAM simulation only (NNCAM+4K). The respective prognostic simulation for the reference climate has a realistic heating rate and temperature structure at the equator but fails at the poles, where temperatures are lower than in the +4K training dataset (Fig.~\ref{fig:S5}\textit{A} and \textit{C}). 

Finally, we train a neural network using half a year of data from the reference and the +4K simulations each, but not the intermediate increments (NNCAM-ref+4k). This version performs well for the extreme climates and also in between (Figs.~\ref{fig:5}\textit{B} and \ref{fig:S5}). Reassuringly, NNCAM-ref+4K is also able to capture important aspects of global warming: an increase in the precipitation extremes (Fig.~\ref{fig:2}\textit{A}) and an amplification and acceleration of the MJO and Kelvin waves (Fig.~\ref{fig:S6}\textit{B}). These sensitivity tests suggest that the neural network is unable to extrapolate much beyond its training climate but can interpolate in between extremes. 

\subsection*{Discussion}
\label{sec:discussion}

In this study we have demonstrated that a deep neural network can learn to represent sub-grid processes in climate models from cloud-resolving model data at a fraction of the computational cost. Freely interacting with the resolved dynamics globally, our deep learning-powered model produces a stable mean climate that is close to its training climate, including precipitation extremes and tropical waves. Moreover, the neural network learned to approximately conserve energy without being told so explicitly. It manages to adapt to new surface forcing patterns but struggles with out-of-sample climates. The ability to interpolate between extremes suggests that short-term, high-resolution simulations which target the edges of the climate space can be used to build a comprehensive training dataset.  Our study shows a potential way for data-driven development of climate and weather models. Opportunities but also challenges abound.

An immediate follow-on task is to extend this methodology to a less idealized model setup and incorporate more complexity in the neural network parameterization. This requires ensuring positivity of water concentrations and stability which we found challenging in first tests. Predicting the condensation rate, which is not readily available in SPCAM, could provide a convenient way to ensure conservation properties. Another intriguing approach would be to predict sub-grid fluxes instead of absolute tendencies. However, computing the flux divergence to obtain the tendencies amplifies any noise produced by the neural network. Future efforts using machine learning parameterizations should systematically address these issues. Additional complexities like topography, aerosols and chemistry will present further challenges but none of those seem insurmountable from our current vantage point.

%Considering additional variables, such as liquid water and ice, should improve the realism of the neural network predictions but raises two challenging issues: First, positive water concentrations have to be ensured in combination with moisture conservation. This is challenging since concentrations are often zero and the network predictions exhibit some scatter. Predicting the condensation rate, which is not readily available in the version of SPCAM we have used, instead of absolute tendencies could provide a convenient way to ensure conservation properties. Second, in preliminary tests, we found that a neural network that includes condensed water suffered from instabilities at individual grid points. Because of the complexity of the neural network and its host model, however, finding the cause for these issues turned out to be non-trivial. Future efforts towards using our methodology for complex Earth System Models must therefore address similar issues. Another intriguing approach would be to predict the convective and radiative sub-grid fluxes instead of the absolute tendencies, which could prove beneficial for ensuring conservation properties. However, computing the flux divergence to obtain the tendencies amplifies any noise produced by the neural network. To ensure stability we therefore refrained from attempting this option but this trade-off would be useful to assess in the near term.  Additional complexities like topography, aerosols and chemistry will present further challenges but none of those seem insurmountable from our current vantage point.

Limitations of our method when confronted with out-of-sample temperatures are related to the traditional problem of overfitting in machine learning---the inability to make accurate predictions for data unseen during training. Convolutional neural networks and regularization techniques are commonly used to fight overfitting.  It may well be possible that a combination of these and novel techniques improves the out-of-sample predictions of a neural network parameterization. Note also that our idealized training climate is much more homogeneous than the real world climate, for instance a lack of the El Ni\~no-Southern Oscillation, which probably exacerbated the generalization issues.

Convolutional and recurrent neural networks could be used to capture spatial and temporal dependencies, such as propagating mesoscale convective systems or convective memory across time steps. Furthermore, generative adversarial networks \cite{Goodfellow2016} could be one promising avenue towards creating a stochastic machine learning parameterization that captures the variability of the training data. Random forests \cite{Breiman2001RandomForests} have also recently been applied to learn and model sub-grid convection in a global climate model \cite{OGorman2018}. Compared to neural networks, they have the advantage that conservation properties are automatically obeyed but suffer from computational limitations.

Recently, it has been argued \cite{Schneider2017a} that machine learning should be used to learn the parameters or parametric functions within a traditional parameterization framework rather than the full parameterization as we have done. Because the known physics are hard-coded this could lead to better generalization capabilities, a reduction of the required data amount and the ability to isolate individual components of the climate system for process studies. On the flip side, it still leaves the burden of heuristically finding the framework equations, which requires splitting a coherent physical system into sub-processes. In this regard, our method of using a single network naturally unifies all sub-grid processes without the need to prescribe interactions.

%Potential advantages of such an approach could be better generalization capabilities and a reduction of the required data amount for training, because more of the physical laws are hard-coded, and the ability to isolate individual components of the climate system for process studies. On the flip side, this still leaves the burden of heuristically finding the framework equations, a process that to some degree always requires making subjective simplifications and splitting one coherent physical system into sub-processes. In this regard, our method of using a single neural network naturally unifies all atmospheric sub-grid processes, a direction in which parameterization development has been heading in recent years \cite{Gentine2013a, Guo2015, Tan2018}. A comparison of our approach with one that prescribes more of the physics would offer important insight into the advantages and disadvantages of each method.

Regardless of the exact type of learned algorithm, once implemented in the prognostic model some biases will be unavoidable. In our current methodology there is no way of tuning after the training stage. We argue, therefore, that an online learning approach, where the machine learning algorithm runs and learns in parallel with a CRM is required for further development. Super-parameterization presents a natural fit for such a technique. For full global CRMs this likely is more technically challenging.
%In our current approach of training the algorithm offline and then implementing it as is, there is no way of tuning after the fact. We argue, therefore, that some kind of online learning procedure is necessary for future use of data-driven parameterizations.  This could involve running the pre-trained machine learning parameterization in parallel with a CRM and updating the algorithm's parameters in its own model climate with the latest high-resolution targets at each time step. Super-parameterization presents a natural fit for such a technique. For full global CRMs this likely is more technically challenging.
%Super-parameterization with its intrinsic scale-separation presents a natural fit for testing such an approach. For global CRMs or locally nested models, how to match up and communicate between the coarse and fine grids could prove to be a more involved technical task.

A grand challenge is how to learn directly from observations---our closest knowledge of the truth---rather than high-resolution simulations which come with their own baggage of tuning and parameterization (turbulence and microphysics) \cite{Schneider2017a}. Complications arise because observations are sparse in time and space and often only of indirect quantities, for example satellite observations. Until data assimilation algorithms for parameter estimation advance, learning from high-resolution simulations seems the more promising route towards tangible progress in sub-grid parameterization.

%The use of such observations to estimate the atmospheric state in numerical weather prediction models using data assimilation has made great progress over the last decades \cite{Bauer2015a} but extensions to learning model parameters are still in their infancy \cite{Simon2015}. Until methods in this area advance, learning from high-resolution simulations might present the more feasible way towards tangible progress in sub-grid parameterizations. 

Our study presents a paradigm shift from the manual design of sub-grid parameterizations to a data-driven approach that leverages the advantages of high-resolution modeling. This general methodology is not limited to the atmosphere but can equally as well be applied to other components of the Earth system and beyond. Challenges must still be overcome, but advances in computing capabilities and deep learning in recent years present novel opportunities that are just beginning to be investigated. We believe that machine learning approaches offer great potential that should be explored in concert with traditional model development.

\subsection*{Acknowledgements}
SR acknowledges funding from the German Research Foundation (DFG) project  SFB/TRR 165 “Waves to Weather”. MP acknowledges funding from the DOE SciDac and Early Career programs DE-SC0012152 and DE-SC0012548 and the NSF programs AGS-1419518 and AGS-1734164. PG acknowledges funding from the NSF programs AGS-1734156 and AGS-1649770, the NASA program NNX14AI36G and the DOE Early Career program DE-SC0014203. Computational resources were provided through the NSF XSEDE allocations TG-ATM120034 and TG-ATM170029. We thank Gaël Reinaudi, David Randall, Galen Yacalis,  Jeremy McGibbon, Chris Bretherton, Phil Rasch, Tapio Schneider, Padhraic Smyth and Eric Nalisnick for helpul conversations during this work.

{\footnotesize \bibliography{library}}
%\printbibliography
\end{multicols}
\clearpage

\section*{Supplement}
\renewcommand{\thefigure}{S\arabic{figure}}
\setcounter{figure}{0}
\renewcommand{\thetable}{S\arabic{table}}
\setcounter{table}{0}

\subsection*{Supplemental Methods}
\label{sec:MM}

\subsubsection*{SPCAM Setup}
The SPCAM model source code along with our modifications, including the neural network implementation, is available at \url{https://gitlab.com/mspritch/spcam3.0-neural-net} (branch: \texttt{nn\_fbp\_engy\_ess}).

We use the Community Atmosphere Model 3.0 \cite{Collins2006} with super-parameterization \cite{Khairoutdinov2001} as our training and reference model. The model has an approximately two-degree horizontal resolution with 30 vertical levels and a 30 minute time step. The embedded two-dimensional cloud resolving models consist of eight 4 km-wide columns oriented meriodinally, as in Ref. \cite{Pritchard2014}. The CRM time step is 20 seconds. Sub-grid turbulence in the CRM is parameterized with a local 1.5-order closure. Each GCM time step the CRM tendencies are applied to the resolved grid. Note that our SPCAM setup does not feed back momentum tendencies from the CRM to the global grid. While these might be important \cite{Moncrieff2017}, our neural network also cannot capture momentum fluxes. Using global CRM data or augmented SP that includes 3D CRM domains with interactive momentum (or 2D SP equipped with a downgradient momentum parameterization after Ref.~\cite{Tulich2015}) would prove beneficial for this purpose, especially towards ocean-coupled simulations in which cumulus friction is known to be important to the equatorial cold tongue/ITCZ nexus \cite{Woelfle2018}. After the SP update, the radiation scheme is called which uses sub-grid cloud information from the CRM. This is followed by a computation of the surface fluxes with a simple bulk scheme and the dynamical core. CTRLCAM uses the default parameterizations which includes the Zhang-McFarlane convection scheme \cite{Zhang1995} and a simple vertical turbulent diffusion scheme. 

The physical parameterization of NNCAM is 20 times faster than SPCAM and 8 times faster than CTRLCAM. This results in a total model speed-up of factor 10 compared to SPCAM and factor 4 compared to CTRLCAM. To generate the best possible training data for the neural network we run the radiation scheme every GCM time step for SPCAM and CTRLCAM. In CTRLCAM, therefore, the radiation scheme is much more computationally expensive than in the standard setup where the radiation scheme is only called every few GCM time steps. 

The sea surface temperatures (SSTs) are prescribed in our aquaplanet setup that follows Ref. \cite{Andersen2012}. The reference state is zonally symmetric with a maximum shifted five degrees to the North of the equator to avoid unstable behaviors observed for equatorially symmetric aquaplanet setups: 
\begin{equation}
    \mathrm{SST}(\phi) = 2 + \frac{27}{2}(2-\zeta-\zeta^2),
\end{equation}
where the SST is given in Celcius, $\phi$ is the latitude in degrees and
\begin{equation}
    \zeta = 
    \begin{cases}
    \sin^2\left(\pi\frac{\phi - 5}{110}\right)       & \quad 5 < \phi \leq 60\\
    \sin^2\left(\pi\frac{\phi - 5}{130}\right)       & \quad -60 \leq \phi < 5\\
    1       & \quad \mathrm{if } |\phi| < 60
  \end{cases}
\end{equation}
Additionally, we run simulations with a globally increased SSTs up to 4K in increments of 1K and a zonally asymmetric run with a wavenumber one perturbation added to the reference SSTs:
\begin{equation}
\begin{split}
    \text{SST}'(\lambda, \phi) = 
            3\cos \left(\frac{\lambda \pi}{180}\right) \cos \left(0.5\pi\frac{(\frac{\phi \pi}{180} -5)}{30}\right)^2 
              \quad  \text{if} \quad -25 \leq \phi \leq 35,
      \end{split}
\end{equation}
where $\lambda$ is longitude in degrees. The sun is in perpetual equinox with a full diurnal cycle. All experiments were started with the same initial conditions and allowed to spin up for a year. The subsequent five years were used for analysis. Training data for the neural network was taken from the second year of the SPCAM simulations.

\subsubsection*{Neural network}
All neural network code is available at \url{https://github.com/raspstephan/CBRAIN-CAM}

We use the Python library Keras \cite{Chollet2015} with the Tensorflow \cite{Tensorflow} backend for all neural network experiments. Our neural network architecture consists of nine fully-connected layers with 256 nodes each. This adds up to a total of 567,361 learnable parameters. The LeakyReLU activation function $\max(0.3x, x)$ resulted in the lowest training losses. The neural network was trained for 18 epochs with a batch size of 1024. The optimizer used was Adam \cite{Kingma2014} with a mean squared error loss function. We started with a learning rate of $1 \times 10^{-3}$ which was divided by five every three epochs. The total training time was on the order of 8 hours on a single Nvidia GTX 1080 graphics processing unit (GPU).

The input variables for the neural network were chosen to mirror the information received by the CRM and radiation scheme but lack the condensed water species and the dynamical tendencies. The latter are applied as a constant forcing during the CRM integration. We found, however, that they did not improve the neural network performance and trimmed the input variables for the sake of simplicity. Another option would be to include the surface flux computation in the network as well. In this option the fluxes are removed from the input and the surface temperature is added. This option yielded similar results but did not allow us to investigate column energy conservation.

The input values are normalized by subtracting each element of the stacked input vector (Table~\ref{tab:vars}) by its mean across samples and then dividing it by the maximum of its range and the standard deviation computed across all levels of the respective physical variable. This is done to avoid dividing by very small values, e.g. for humidity in the upper levels, which can cause the input values to become very large if the neural network predicts noisy tendencies. For the outputs, the heating and moistening rates are brought to the same order of magnitude by converting them to W kg$^{-1}$ . The radiative fluxes and precipitation were normalized to be on the same order of magnitude as the heating and moistening rates (see Table~\ref{tab:vars} for multiplication factors). The magnitude of the output values determines their importance in the loss function. In our quadratic loss function differences are highlighted even further. Making sure that no single value dominates the loss is important to get a consistent prediction quality. For a reasonable range (factor five) around our normalization values the results are largely unaffected, however. 

\begin{figure}[h!]
	\centering
	\includegraphics[width=1\linewidth]{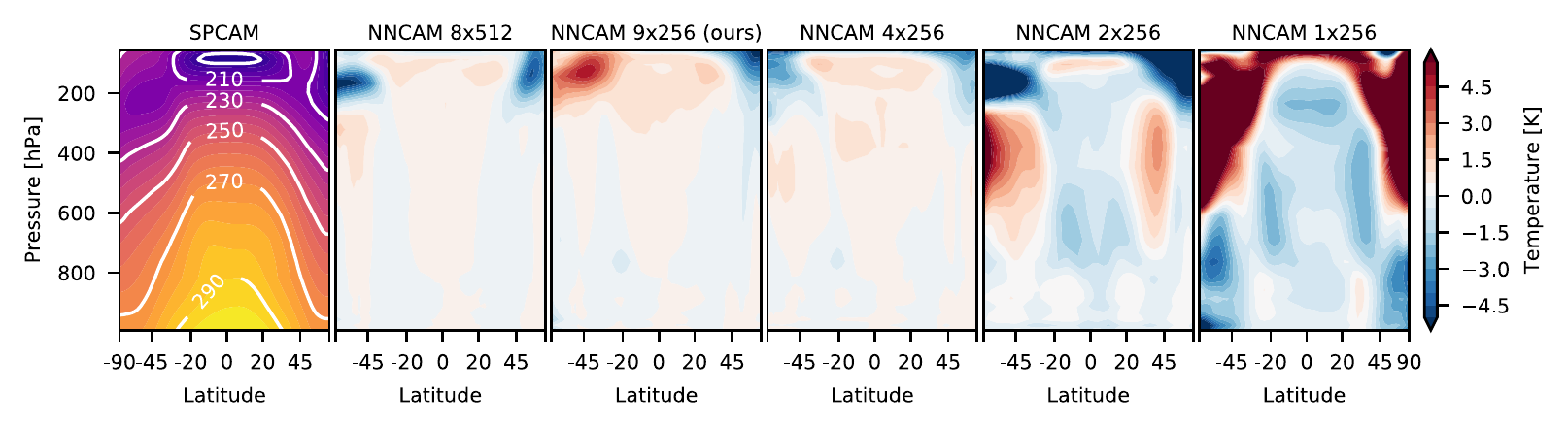}
	\caption{All figures show longitudinal and five year-temporal averages as in Fig.~1. Zonally and temporally averaged temperature relative to SPCAM for different network configurations (Number of hidden layers x Nodes per hidden layer). 8x512 corresponds to the network in Ref. \cite{Gentine2018}.}
	\label{fig:S1}
\end{figure}

Deep neural networks appear to be essential to achieve a stable and realistic prognostic implementation. Similar to other studies which used shallow neural networks \cite{Krasnopolsky2013, Brenowitz2018} we encountered unstable modes and unrealistic artifacts for networks with two or one hidden layers (Fig.~\ref{fig:S1}). A four layer network was the minimal complexity to provide good results for our configuration. Adding further layers shows little correlation between training skill and prognostic performance. We chose our network design to lie well withing the range of stable network configurations.

\begin{table}[h!]
\centering
\caption{Table showing input and output variables and their number of vertical levels $N_z$. For the output variables the normalization factors are also listed. $C_p$ is the specific heat of air. $L_v$ is the latent heat of vaporization. }
\label{tab:vars}
\small
\begin{tabular}{lll|llll}
\textbf{Input variables}     & \textbf{Unit}   & $\mathbf{N_z}$ & \textbf{Output variables}& \textbf{Unit}                 & $\mathbf{N_z}$ & \textbf{Normalization} \\ \hline \hline
Temperature &K      & 30     & Heating rate  $\Delta T_{\mathrm{phy}}$& K s$^{-1}$ & 30     & $C_p$          \\
Humidity   &kg kg$^{-1}$             & 30     & Moistening rate      $\Delta Q_{\mathrm{phy}}$    &kg kg$^{-1}$ s$^{-1}$              & 30     & $L_v$        \\
Meridional wind  &m s$^{-1}$       & 30     & Shortwave flux at TOA &W m$^{-2}$ & 1      & $10^{-3}$             \\
Surface pressure & Pa       & 1      & Shortwave flux at surface     &  W m$^{-2}$        & 1      & $10^{-3}$              \\
Incoming solar radiation& W m$^{-2}$& 1      & Longwave flux at TOA & W m$^{-2}$ & 1      & $10^{-3}$              \\
Sensible heat flux  &W m$^{-2}$     & 1      & Longwave flux at surface& W m$^{-2}$              & 1      & $10^{-3}$              \\
Latent heat flux  & W m$^{-2}$      & 1      & Precipitation          &  kg m$^{-2}$ d$^{-1}$               & 1      &  $2 \times 10^{-2}$               \\ \hline
Size of stacked vectors&  & 94     &                                  &       & 65     &              
\end{tabular}
\end{table}

\begin{figure}[h!]
    \centering
    \includegraphics[width=1\linewidth]{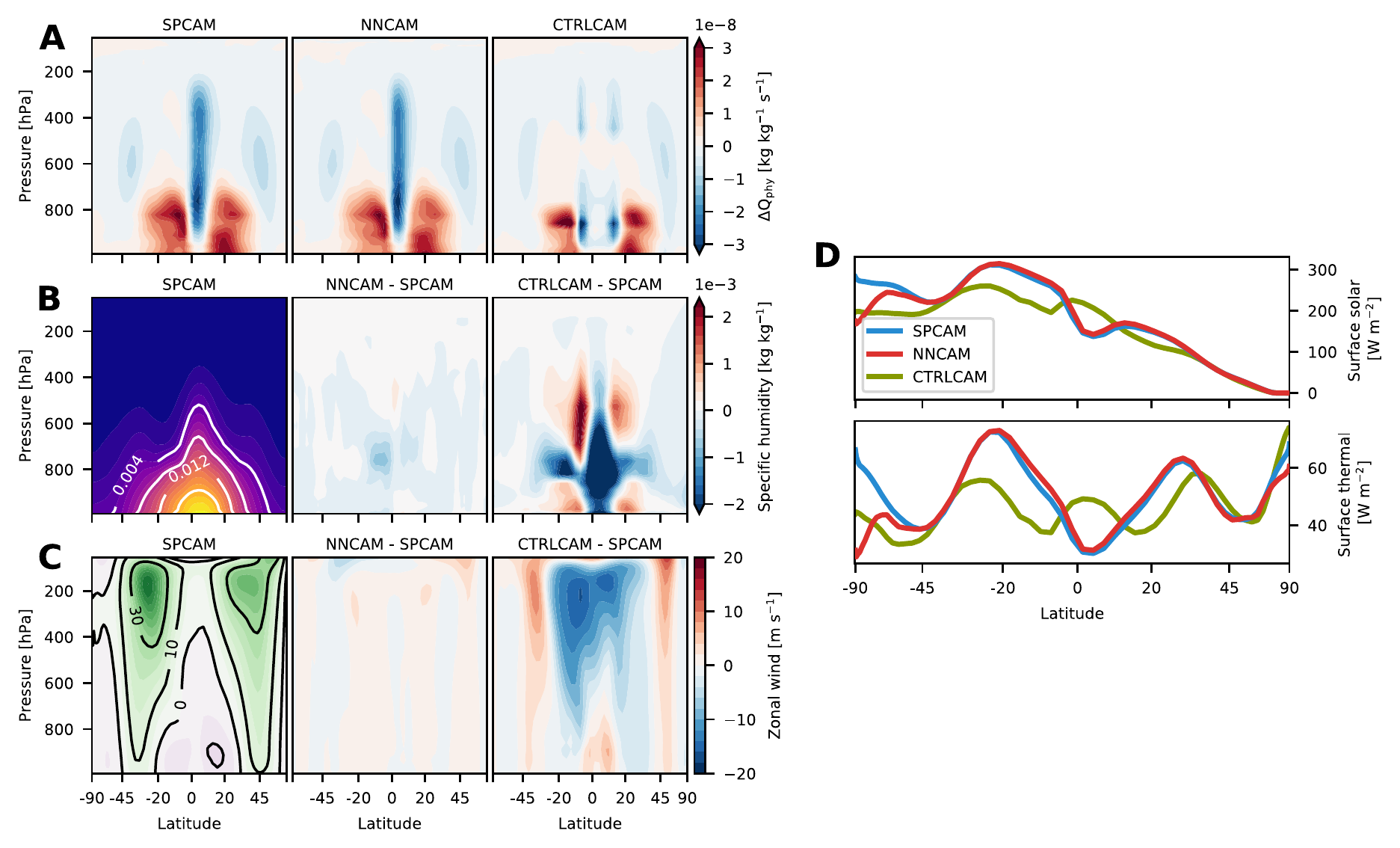}
    \caption{(A) Mean convective sub-grid moistening rates $\Delta Q_{\mathrm{phy}}$. (B) Mean specific humidity $Q$ and (C) zonal wind $V$ of SPCAM and biases of NNCAM and CTRLCAM relative to SPCAM. (D) Mean shortwave (solar) and longwave (thermal) net fluxes at the surface. The latitude axis is area-weighted.}
    \label{fig:S2}
\end{figure}

\begin{figure}[h!]
    \centering
    \includegraphics[width=0.6\linewidth]{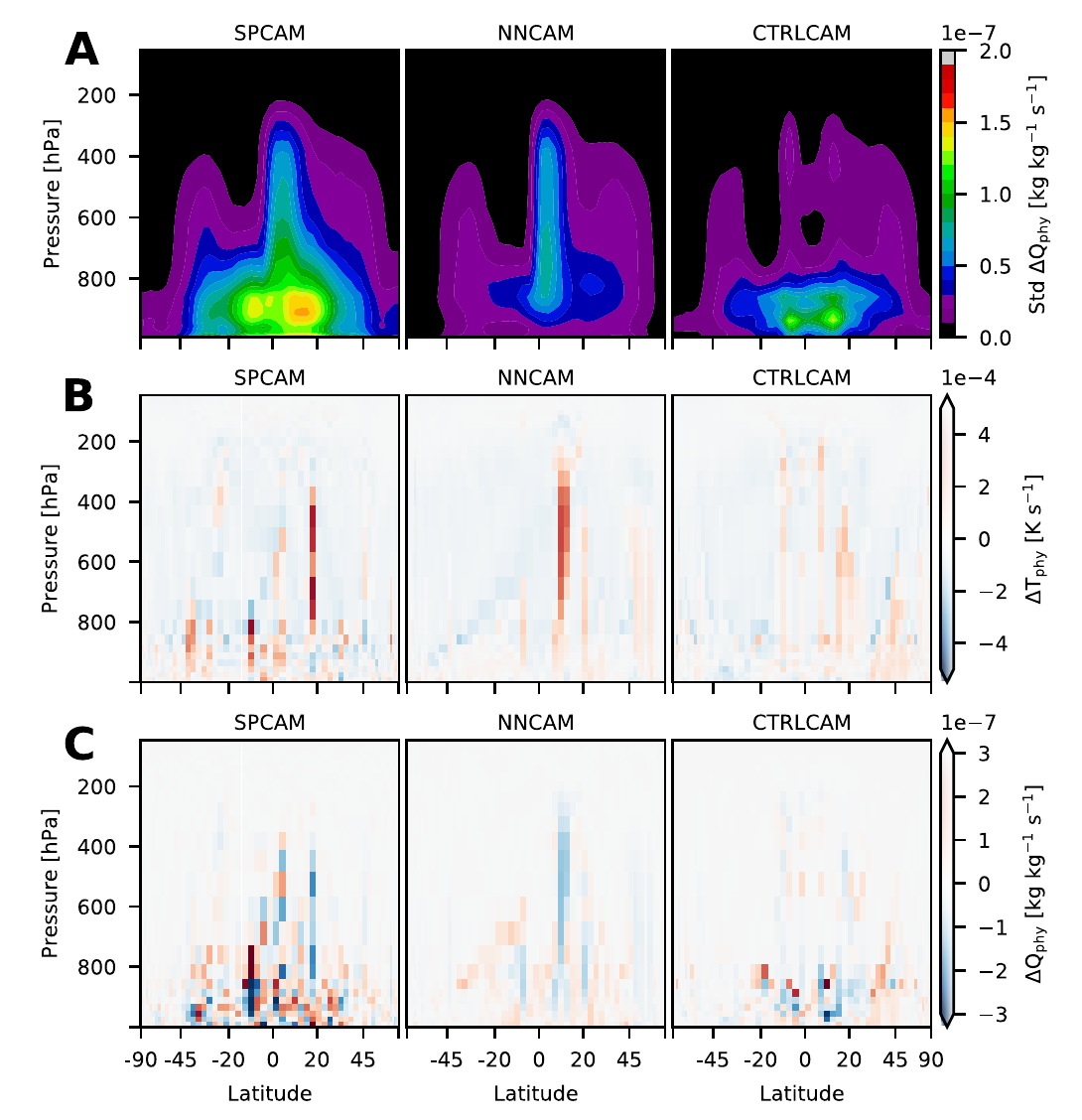}
    \caption{(A) Zonally averaged temporal standard deviation of the convective sub-grid moistening rate $\Delta Q_{\mathrm{phy}}$. (B, C) Snapshots of heating $\Delta T_{\mathrm{phy}}$ and moistening rate $\Delta Q_{\mathrm{phy}}$. Note that these are taken from the free model simulations and should, therefore, not correspond one-to-one between the experiments.}
    \label{fig:S3}
\end{figure}

\begin{figure}[h!]
    \centering
    \includegraphics[width=0.7\linewidth]{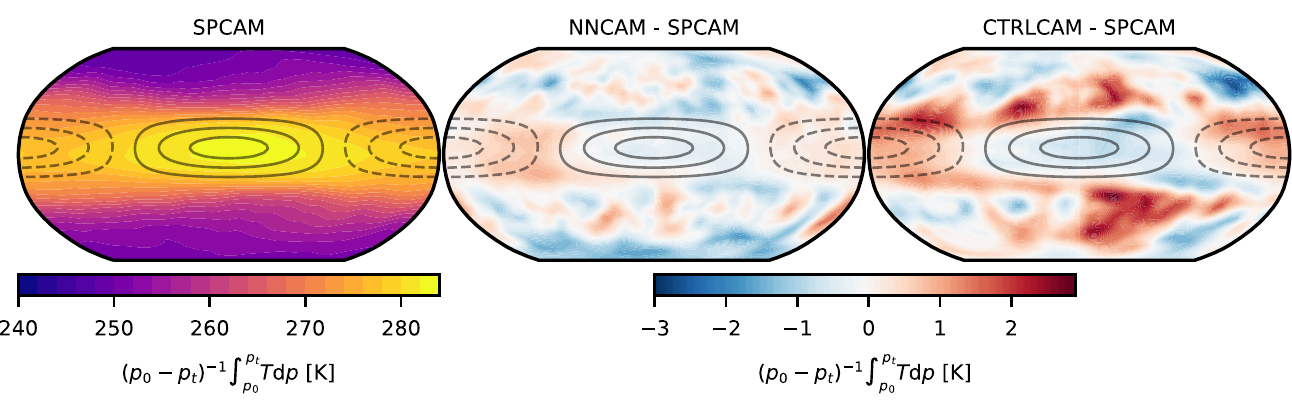}
    \caption{Mass-weighted temperature integrated over the troposphere from $p_0 = 1000$ hPa to $p_t = 380$ hPa for SPCAM reference and differences of NNCAM and CTRLCAM with respect to reference for zonally perturbed simulations.}
    \label{fig:S4}
\end{figure}

\begin{figure}[h!]
    \centering
    \includegraphics[width=0.88\linewidth]{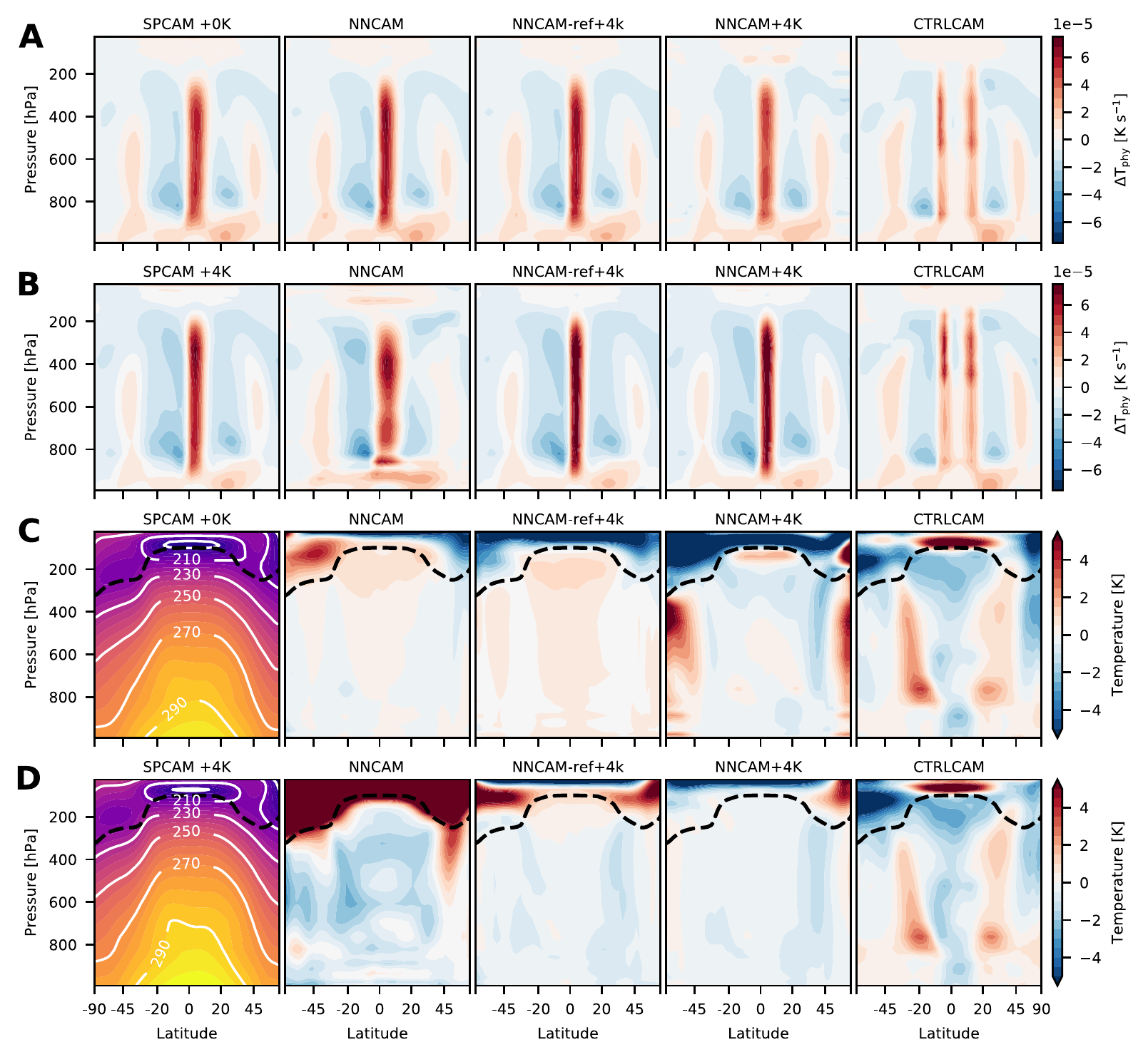}
    \caption{Zonally and temporally averaged (A, B) heating rate and (C, D) temperature relative to SPCAM. Panels A and C show reference SSTs while panels B and D show global 4 K perturbation. Temperature panels show SPCAM reference and differences to reference for several experiments described in the text. }
    \label{fig:S5}
\end{figure}

\begin{figure}[h!]
    \centering
    \includegraphics[width=0.6\linewidth]{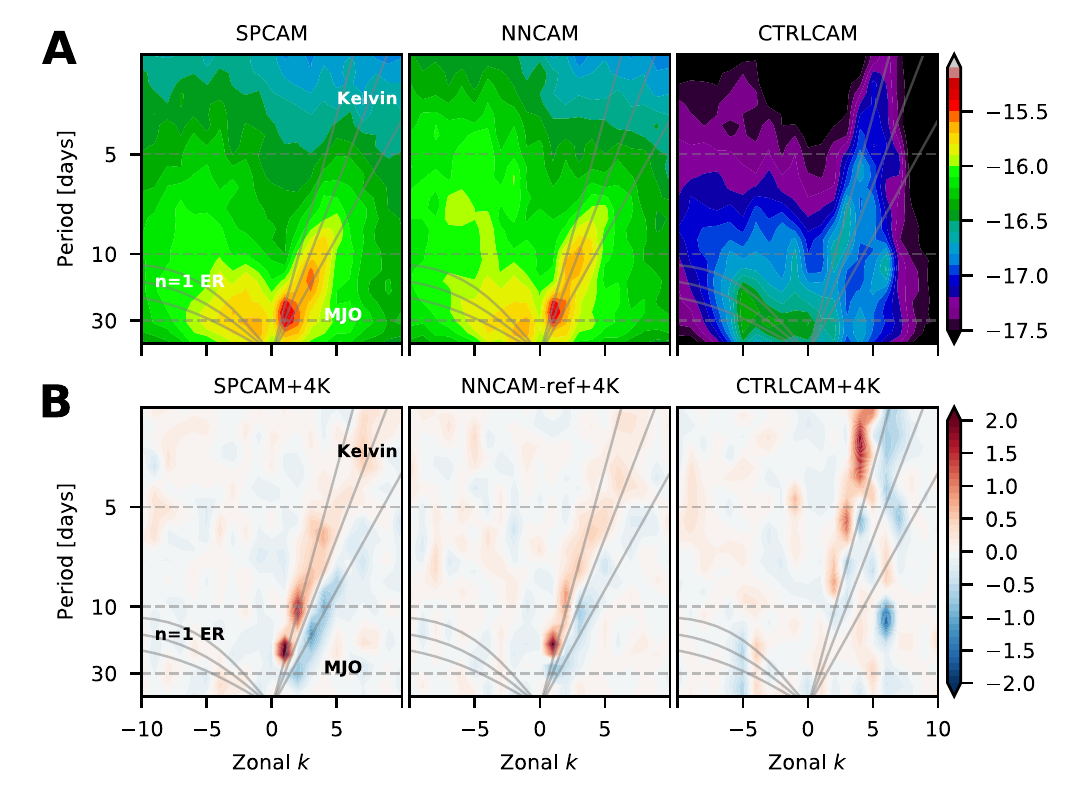}
    \caption{(A) Space-time spectrum of the equatorially symmetric component of 15S-15N daily precipitation anomalies. As in Fig. 1b of Ref. \cite{Wheeler1999}. (B) Space-time spectrum of the equatorially symmetric component of 15S-15N daily precipitation anomalies divided by background spectrum. As in Fig. 3b of Ref. \cite{Wheeler1999}. Figure shows +4K SST minus reference SST. Negative (positive) values denote westward (eastward) traveling waves.}
    \label{fig:S6}
\end{figure}

\end{document}